\documentclass[preprint,showpacs,preprintnumbers,amsmath,amssymb,graphicx]{revtex4}
\usepackage{amsmath,amsfonts,amssymb}
\usepackage{epsfig}
\def\bit{\begin{itemize}}
\def\eit{\end{itemize}}
\def\ben{\begin{enumerate}}
\def\een{\end{enumerate}}
\def\bed{\begin{description}}
\def\eed{\end{description}}

\def\lsim{\raise0.3ex\hbox{$<$\kern-0.75em\raise-1.1ex\hbox{$\sim$}}}
\def\gsim{\raise0.3ex\hbox{$>$\kern-0.75em\raise-1.1ex\hbox{$\sim$}}}

\let\jnfont=\rm
\def\NPB#1,{{\jnfont Nucl.\ Phys.\ B }{\bf #1},}
\def\PLB#1,{{\jnfont Phys.\ Lett.\ B }{\bf #1},}
\def\EPJC#1,{{\jnfont Eur.\ Phys.\ Jour.\ C }{\bf #1},}
\def\PRD#1,{{\jnfont Phys.\ Rev.\ D }{\bf #1},}
\def\PRL#1,{{\jnfont Phys.\ Rev.\ Lett.\ }{\bf #1},}
\def\MPLA#1,{{\jnfont Mod.\ Phys.\ Lett.\ A }{\bf #1},}
\def\JPG#1,{{\jnfont J.\ Phys.\ G}{\bf #1},}
\def\CTP#1,{{\jnfont Commun.\ Theor.\ Phys.\ }{\bf #1},}
\def\JHEP#1,{{\jnfont JHEP \ }{\bf #1},}
\def\NPPS#1,{{\jnfont Nucl.\ Phys.\ Proc.\ Suppl.\ }{\bf #1},}

\def\beq{\begin{equation}}
\def\eeq{\end{equation}}
\def\bea{\begin{eqnarray}}
\def\eea{\end{eqnarray}}
\newcommand{\ba}{\begin{array}}
\newcommand{\ea}{\end{array}}

\begin{document}

\title{ Constraints of dark matter direct detection experiments on the MSSM
        and implications for LHC Higgs searches}

%\title{Constraining the MSSM with the direct detection experiments of the dark matter
%       and its implication in Higgs search }

\author{\ \\[1mm]
Junjie Cao$^1$, Ken-ichi Hikasa$^2$, Wenyu Wang$^3$, Jin Min Yang$^4$, Li-Xin Yu$^4$ \\ ~
}

\affiliation{
$^1$ Department of Physics, Henan Normal University, Xinxiang 453007, China\\
$^2$ Department of Physics, Tohoku University, Sendai 980-8578, Japan \\
$^3$ Institute of Theoretical Physics, College of Applied Science,
     Beijing University of Technology, Beijing 100124, China\\
$^4$ Institute of Theoretical Physics, Academia Sinica,
              Beijing 100190, China
\vspace*{.1cm} }

\begin{abstract}
Assuming the lightest neutralino solely composes the cosmic dark
matter, we examine the constraints of the CDMS-II and XENON100 dark
matter direct searches on the parameter space of the MSSM Higgs
sector. We find that the current CDMS-II/XENON100 limits can exclude
some of the parameter space which survive the constraints from the
dark matter relic density and various collider experiments. We also
find that in the currently allowed parameter space, the charged
Higgs boson is hardly accessible at the LHC for an integrated
luminosity of 30 fb$^{-1}$, while the neutral non-SM Higgs bosons
($H$,$A$) may be accessible in some allowed region characterized by
a large $\mu$.
The future XENON100 (6000 kg-days exposure) will significantly
tighten the parameter space in case of nonobservation of dark matter,
further shrinking the likelihood of discovering the non-SM Higgs bosons
at the LHC.
\end{abstract}
\maketitle

{\em Introduction:~} The existence of non-baryonic cold dark matter (DM)
has been established by cosmological observations \cite{Dunkley}. Weakly
interacting massive particles (WIMPs) are the natural candidates of DM,
among which the lightest neutralino $\chi^0_1$ in the
minimal supersymmetric standard model (MSSM) has been most extensively
studied \cite{mssm}.

The most convincing detection for the neutralino DM is the
underground direct detection experiments like CDMS and XENON, which
search for neutralino-nucleon ($\chi N$) scattering in a
low-background circumstance \cite{cdms,xenon100-1}. Recently, both
CDMS-II and XENON100 reported their search results
\cite{cdms,xenon100-1}, which immediately stimulated some
theoretical works \cite{Hisano,Asano}. On the theoretical side,
great efforts have been paid to improve the accuracy of the
prediction for the $\chi N$ scattering rate
\cite{Drees,Djoudi}. For example, it has long been known that the
hadronic uncertainty, especially the strange quark content in a
nucleon, can affect the rate by almost one order of magnitude, and
is therefore impacting significantly the interpretation of the
experimental searches for DM \cite{Ellis}. This problem was recently
better solved by lattice simulation and it is found that the strange
quark content is much smaller than previously thought, which leads
to a significant suppression on the uncertainty \cite{lattice}.

In light of the above experimental and theoretical progress in DM
study, we in this work re-investigate the $\chi N$
scattering and use the recent CDMS-II/XENON limits to constrain the
MSSM parameter space. Note that unlike most recent studies which try
to explain the two possible DM events reported by CDMS-II, we use
the CDMS-II $90\%$ upper limit on the spin-independent (SI)
$\chi N$ scattering cross section. In addition to the direct
detection limits from CDMS-II and XENON100, we also consider the
constraints from the DM relic density and various collider
experiments. We will first perform a scan over the MSSM parameter
space by considering these constraints. Then we investigate the
$\chi N$ scattering in the surviving parameter space to
demonstrate the further constraints of CDMS-II/XENON on it.
Given the extreme importance of Higgs search at the LHC and
the strong correlation between the $\chi N$ scattering and
the Higgs sector, our study will be focused on the MSSM Higgs
sector.

{\em\boldmath $\chi N$ scattering in the MSSM:~} For the sensitivity
in current DM direct detection experiments, it is sufficient to
consider only the SI interactions between $\chi^0_1$ and
nucleon (denoted by $f_p$ for proton and $f_n$ for neutron
\cite{susy-dm-review}) in calculating the scattering rate. In the
MSSM, these interactions are induced  by exchanging squarks or
neutral Higgs bosons at tree level \cite{Drees,susy-dm-review}. For
moderately light Higgs bosons, the latter contribution is dominant
and $f_p$ is approximated by \cite{susy-dm-review} (similarly for
$f_n$)
\begin{equation}
 \begin{split}
    f_{p}
  \simeq
%%%& \sum f_q^{H} \langle p  | \bar q q | p  \rangle \\ =&
\sum_{q=u, d, s} \frac{f_q^{H}}{m_q} m_p f_{T_q}^{(p)}
    + \frac{2}{27}f_{T_G} \sum_{q=c, b, t} \frac{f_q^{H}}{m_q} m_p,
 \end{split}     \label{2b}
\end{equation}
where $f_{Tq}^{(p)}$ denotes the fraction of $m_p$ (proton mass)
from the light quark $q$ while
$f_{T_G}=1-\sum_{u,d,s}f_{T_q}^{(p)}$ is the heavy quark
contribution through gluon exchange. $f_q^{H}$ is the
coefficient of the effective scalar operator given
by \cite{susy-dm-review}
\begin{equation}
    f_q^{H} = m_q \frac{g_2^2}{4 m_W}
    \biggl( \frac{C_{h \chi \chi}  C_{hqq}}{m_{h}^2}
    + \frac{C_{H  \chi \chi}  C_{Hqq}}{m_{H}^2}     \biggr),
    \label{Higgs-contr}
\end{equation}
with $C$ standing for the corresponding Yukawa couplings. The
$\chi^0_1$--nucleus scattering rate is then given by
\cite{susy-dm-review}
\begin{equation}
    \sigma^{SI} = \frac{4}{\pi}
    \left( \frac{m_{\chi^0} m_T}{m_{\chi^0} + m_T} \right)^2
    \times \bigl( n_p f_p + n_n f_n \bigr)^2,
\end{equation}
where $m_T$ is the mass of target nucleus and $n_p (n_n)$ is the number of
proton (neutron) in the target nucleus.

From the above formulas we can infer in which situation the
scattering cross section is large. Eq.(\ref{Higgs-contr}) indicates
that this occurs only when $C_{S \chi \chi}$ and/or $C_{Sqq}$ ($S$
stands for a Higgs boson) get enhanced. Since the potential
enhancement of $C_{Hdd}$ by $\tan \beta$ is well known, we here only
analysis the behavior of $C_{S \chi \chi}$ with the variation of
SUSY parameters. For a bino-like $\chi^0_1$ encountered in this
work, this coupling is generated through the bino-higgsino mixing
\cite{mssm}, so a large $C_{S \chi \chi}$ needs a large mixing,
which means a small $\mu$. To make this statement clearer, one may
consider the limit of $M_1 \ll M_2, \mu$ with $M_1$, $M_2$ and $\mu$
denoting respectively the masses of bino, wino and higgsino. After
diagonalizing the neutralino mass matrix perturbatively, one can get
\cite{Hisano}
\begin{equation}
 \begin{split}
    &C_{h \chi \chi} \simeq
  \frac{m_Z \sin \theta_W \tan \theta_W}{M_1^2 - \mu^2}
    \bigl[ M_1 + \mu \sin2 \beta \bigr], \\
  &C_{H \chi \chi} \simeq
   - \frac{m_Z \sin \theta_W \tan \theta_W}{M_1^2 - \mu^2}
    \mu \cos2\beta .
 \end{split}     \label{2d}
\end{equation}
So both couplings become large when $\mu$ approaches downward to
$M_1$.

In our numerical calculations for the scattering rate, we considered
all the contributions known so far, including the QCD correction,
SUSY-QCD correction \cite{Djoudi} as well as the contribution from
high dimensional operators \cite{susy-dm-review}. Note that the
SUSY-QCD corrections are not negligible because they may sizably
reduce the scattering rate by suppressing $C_{Sqq}$
\cite{Djoudi,Carena}. In our calculations we take $f_{T_u}^{(p)} =
0.023$, $f_{T_d}^{(p)} = 0.034$, $f_{T_u}^{(n)} = 0.019$,
$f_{T_d}^{(n)} = 0.041$ and $f_{T_s}^{(p)} = f_{T_s}^{(n)} = 0.020$.
Note the value of $f_{T_s}$ we choose is much smaller than that
taken in most previous studies. This small value comes from the
recent lattice simulation \cite{lattice}, and it can reduce the
scattering rate significantly.

{\em Numerical results:~} We make some assumptions to reduce the
number of free parameters before our scan. First, we note that the
first two generation squarks may be heavier than about 400 GeV from
the Tevatron experiments \cite{squark}, and thus their effects on
the scattering should be unimportant in the presence of light Higgs
bosons. So, for the first two generation squarks we fix the soft
masses and the trilinear parameters to be 1 TeV. We checked our
conclusion are not affected by such specific choice. Second, since
the third generation squarks affect the Higgs sector significantly,
we let free all the relevant soft parameters. But to simplify our
analysis, we assume $m_{\tilde{D}_3} =m_{\tilde{U}_3}$ and $A_b =
A_t$, which is well motivated by the mSUGRA model with large $\tan
\beta$. Third, we note that, although the slepton masses do not
directly affect the $\chi N$ scattering rate, they can
affect the allowed range of $\tan \beta$ via the muon $g-2$. In
order to avoid a tight constraint on $\tan \beta$, we assume a
universal soft parameter $m_{\tilde{\ell}}$ and vary it in our scan.
Finally, we use the grand unification relation $3 M_1/5
\alpha_1=M_2/\alpha_2=M_3/\alpha_3$ for the gaugino masses.

With the above assumptions, the free parameters remained are scanned
in the ranges: $ 1 \leq \tan \beta \leq 80 $, $ 80 {\rm~GeV} \leq
m_A \leq 300 {\rm~GeV} $, $ 30 {\rm~GeV} \leq M_1 \leq 500 {\rm~GeV}
$, $ 100 {\rm~GeV} \leq \mu, m_{\tilde{\ell}}, m_{\tilde{Q}_3},
m_{\tilde{U}_3} \leq 1 {\rm~TeV} $ and $ - 3 {\rm~TeV} < A_t \leq  3
{\rm~TeV} $. In our scan,  we consider the following constraints
like done in \cite{Cao1}: (1) Direct bounds on sparticle and Higgs
masses from LEP and Tevatron experiments. (2) LEP II search for
Higgs bosons, which includes various channels of Higgs boson
productions. (3) LEP I and LEP II constraints on the productions of
neutralinos and charginos. (4) Constraints ($2 \sigma$) from
precision electroweak observables plus $R_b$ \cite{Cao}, and also
the constraints from B-physics observables such as $B \to X_s
\gamma$, $B_s \to \mu^+\mu^-$, $B^+ \to \tau^+ \nu$, and the mass
difference $\Delta M_d$ and $\Delta M_s$. (5) The muon $g-2$
constraint \cite{Davier} (we require the MSSM contribution to
explain the deviation at $2 \sigma$ level). (6) We require
$\chi_1^0$ to account for the WMAP measured dark matter
relic density at $2\sigma$ level \cite{Dunkley}. The samples
surviving the above constraints will be input for the calculation of
the $\chi N$ scattering rate. Note that  most of the above
constraints have been encoded in NMSSMTools \cite{NMSSMTools}. We
extended the code to the MSSM case, especially we wrote the code for
the $\chi N$ scattering rate to improve the scan efficiency.
%%%%fig.1 %%%%%%%%%%%%%%%%%%%%%%%%%%%%%%%%%%%%%%%%%%%
\begin{figure}[htbp]
\includegraphics[width=13cm]{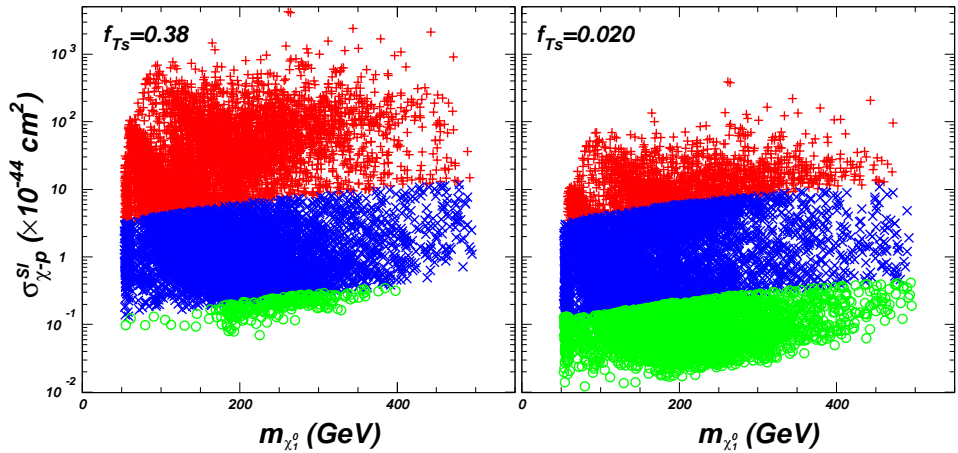}
\vspace{-0.5cm}
\caption{The scatter plots for the spin-independent elastic cross
section of $\chi N$ scattering under the constraints of dark matter
relic density ($2\sigma$) and various collider experiments.
The `$+$' points (red) are excluded
by CDMS II and XENON100 (90\%\ C.L.) limits,
the `$\times$' (blue) would be further excluded by  XENON100 (6000 kg-days)
in case of nonobservation, and the `$\circ$' (green) are beyond the
XENON100 (6000 kg days) sensitivity.}
\label{fig1}
\end{figure}
%%%%%%%%%%%%%%%%%%%%%%%%%%%%%%%%%%%%%%%%%%%%%%%%%%%%%%%

To show the sensitivity of the $\chi N$ scattering rate to the
value of $f_{Ts}$,
we plot the surviving samples on the plane of the scattering rate
versus the neutralino mass with the new lattice value $f_{T_s}=0.02$
and with the old value $f_{Ts}=0.38$ (corresponding to
$\Sigma_{\pi N} \simeq 64$ MeV in \cite{Ellis}).
One can see that the new lattice value of $f_{T_s}$
gives a much lower scattering rate. In our following results
we fixed $f_{T_s}=0.02$.

Our scan samples are $2 \times 10^{11}$ random points over
the parameter space, and after considering the constraints,
about $6 \times 10^7$ samples can survive.  We find that for some
survived samples the $\chi N$ scattering rate can be as large as
$10^{-42}$ cm$^2$, which is far above the current CDMS II/XENON
limits \cite{cdms}.
Requiring the scattering rate not exceed the current CDMS II/XENON
bounds, we find that about $33\%$ ($61\%$) of the survived samples are ruled
out for $m_A \leq 300$ GeV (200 GeV).
Further, if the future XENON100
with 6000 kg-days exposure \cite{xenon100-2} gives null dark matter
results, then $90.5\%$ ($99.5\%$) of the survived samples
will be ruled out for $m_A \leq 300$ GeV (200 GeV).
So the dark matter direct detection experiments
are highly complementary to collider experiments in testing the MSSM.

%%%%fig.2 %%%%%%%%%%%%%%%%%%%%%%%%%%%%%%%%%%%%%%%%%%%
\begin{figure}[htb]
\includegraphics[width=15cm]{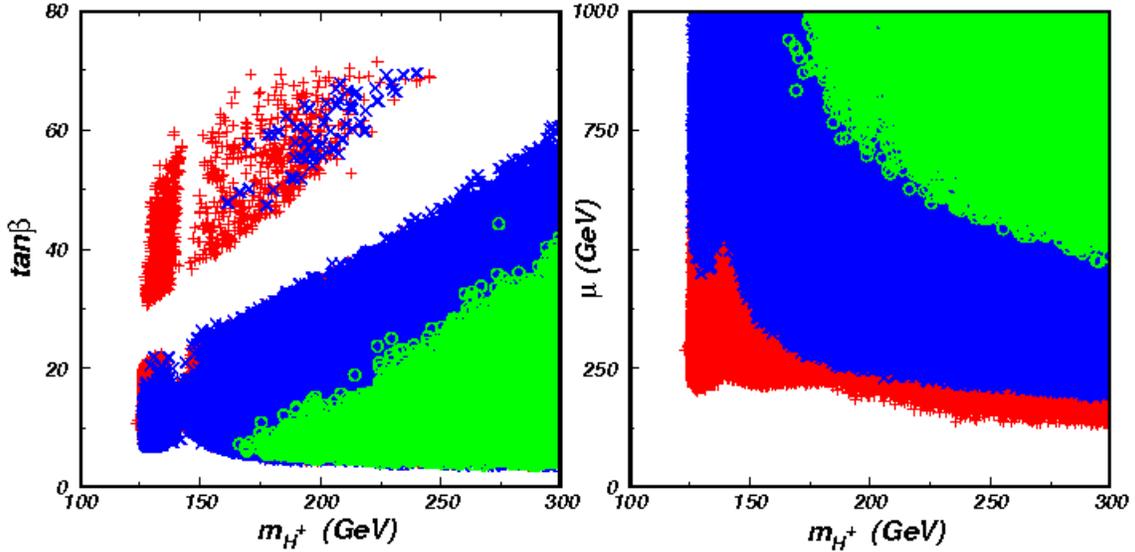}
\vspace{-0.8cm}
\caption{Same as Fig.~\ref{fig1}, but showing $\tan \beta$ and $\mu$ versus $m_{H^+}$
         for $f_{T_s}=0.02$.}
\label{fig2}
\end{figure}
%%%%%%%%%%%%%%%%%%%%%%%%%%%%%%%%%%%%%%%%%%%%%%%%%%%%%%%
In Fig.~\ref{fig2} the surviving samples are projected for $\tan
\beta$ and  $\mu$ versus the charge Higgs boson mass $m_{H^+}$. We
see that the regions excluded by the CDMS-II/XENON limits are
characterized by large $\tan \beta$ and small $\mu$. The future
XENON100 (6000 kg-days) can further shrink the currently allowed
regions, and in particular set a bound $m_{H^+} \gtrsim 165$ GeV.
Around this lower bound, the value of $\mu$ is quite large
($\simeq 1$ TeV) and so the MSSM has a fine-tuning due to the
relation $m_{H^+}^2 > m_A^2$ and $m_A^2= m_{h_u}^2 + m_{h_d}^2 + 2
\mu^2 $ \cite{mssm}.

\begin{figure}[htb]
\includegraphics[width=15cm]{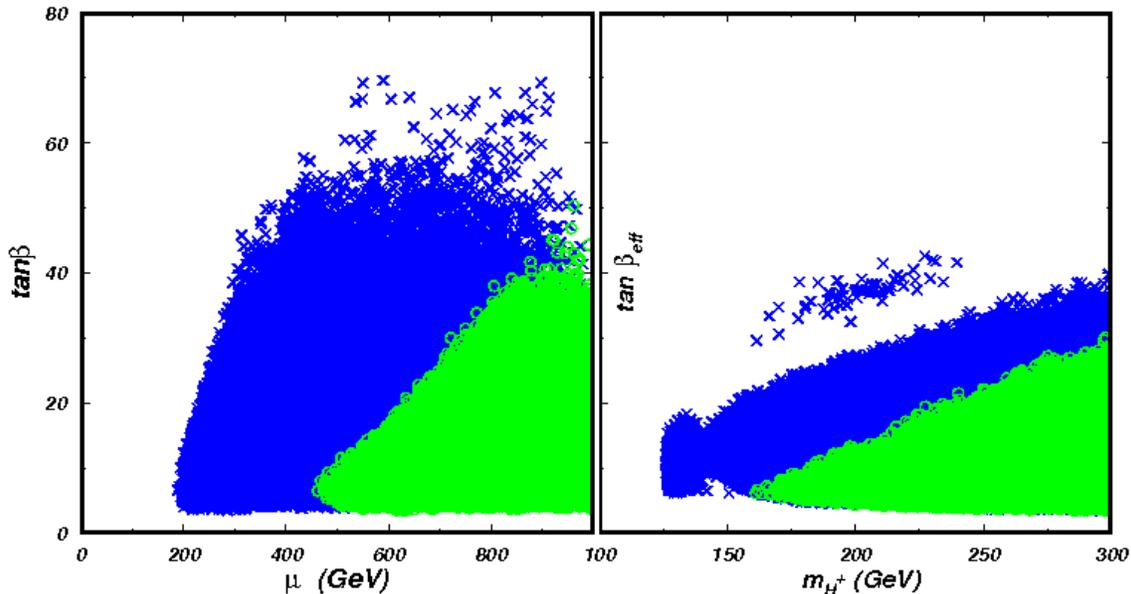}
\vspace{-0.8cm}
\caption{Same as Fig.~\ref{fig2}, but showing  $\tan \beta$ versus $\mu$
and $\tan \beta_{\rm eff}$ versus $m_{H^+}$.}
\label{fig3}
\end{figure}

In our following discussions, we only focus on the samples that
satisfy the current CDMS-II/XENON limits. In Fig.~\ref{fig3} we
project the surviving samples in the planes of $\tan \beta$--$\mu$ and
$\tan \beta_{\rm eff}$--$m_{H^+}$. Here $\tan \beta_{\rm eff}\equiv \tan
\beta/(1+ \Delta_b)$ with $\Delta_b$ denoting the SUSY radiative
corrections to bottom quark mass\cite{Carena}. As expected, large
$\tan \beta$ must be accompanied by large $\mu$ to suppress the the
scattering rate, and this tendency becomes more apparent for the
samples further satisfying the future XENON100 limit. We note that
in this case, i.e. large $\tan \beta$ along with large $\mu$,
$\Delta_b$ should be large\cite{Carena} so that $\tan \beta_{\rm eff}$
is significantly smaller than $\tan \beta$. This speculation is
verified by Fig.~\ref{fig3} and also by our results for $\Delta_b$,
which show $\Delta_b$ larger than $30\%$ for $\tan \beta \geq 40$.
%%%%fig.3 %%%%%%%%%%%%%%%%%%%%%%%%%%%%%%%%%%%%%%%%%%%
%%%%%%%%%%%%%%%%%%%%%%%%%%%%%%%%%%%%%%%%%%%%%%%%%%%%%%%

%%%%fig.4 %%%%%%%%%%%%%%%%%%%%%%%%%%%%%%%%%%%%%%%%%%%
\begin{figure}[htb]
\includegraphics[width=15cm]{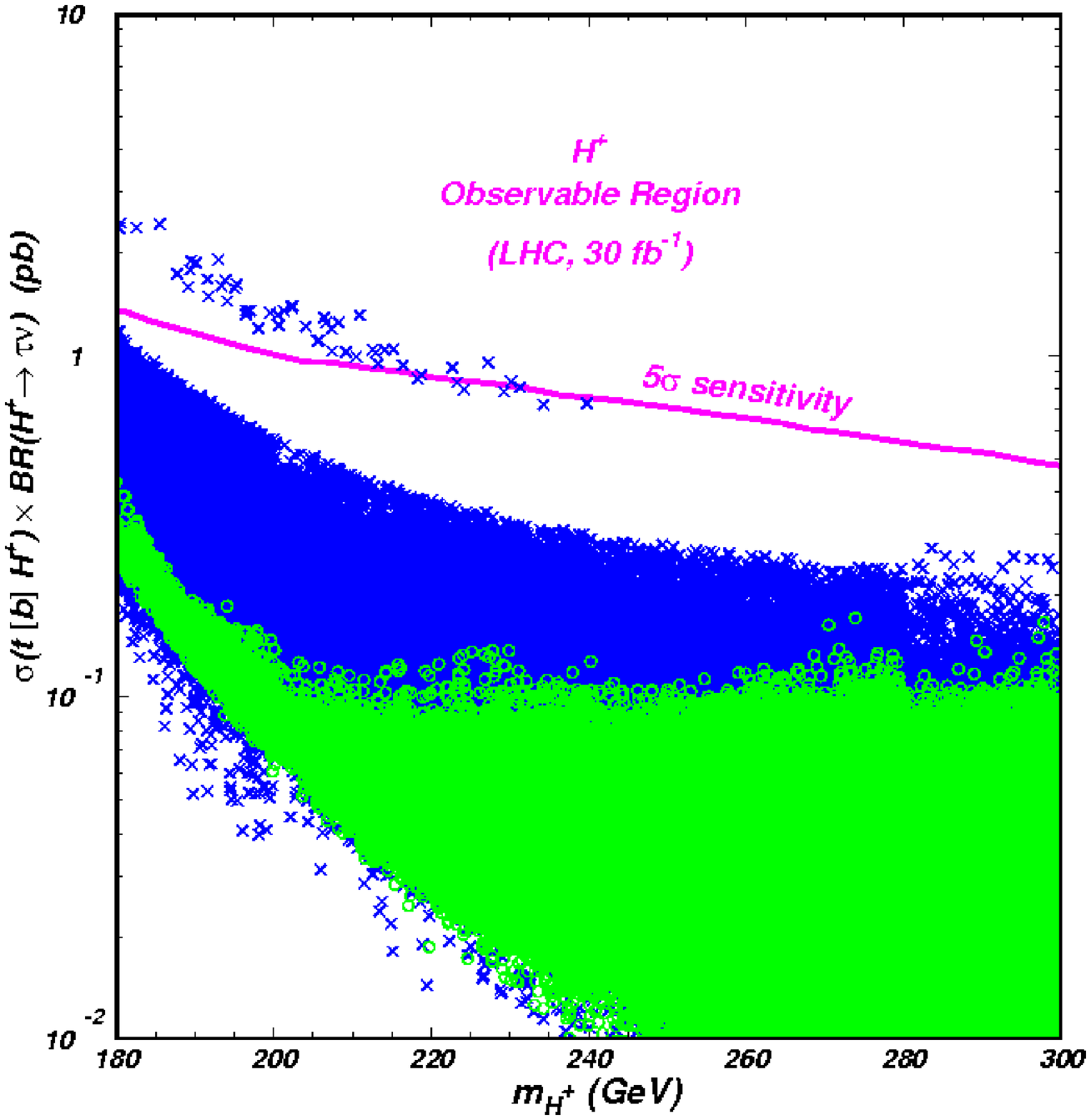}
\vspace{-0.8cm} \caption{Same as Fig.~\ref{fig3},
         but showing the LHC search sensitivity for the charged Higgs boson \cite{ATLAS}.
Here the LHC sensitivity curve obtained by the ATLAS collaboration
\cite{ATLAS} corresponds to the $5 \sigma$ discovery level, while
the exclusion limit from the dark matter direct detection experiments
is at $90\%$ C.L.} \label{fig4}
\end{figure}
%%%%%%%%%%%%%%%%%%%%%%%%%%%%%%%%%%%%%%%%%%%%%%%

{\em Implication for LHC Higgs searches:~} Above results showed that
the CDMS-II/XENON limits have set upper bounds on $\tan \beta_{\rm
eff}$. Since the LHC search for non-SM Higgs boson usually needs a
large $\tan \beta_{\rm eff}$ to enhance the signal rate
\cite{ATLAS,CMS-curve,Horvat}, such upper bounds on $\tan \beta_{\rm
eff}$ may have important implication on LHC search for non-SM Higgs
bosons.

We first consider the LHC search for the charged Higgs boson, which,
for the charged Higgs heavier than top quark, mainly utilizes the
channel $gg/gb \to t[b]H^+$ with $H^+$ subsequently decaying to
$\tau^+ \nu_\tau$ \cite{ATLAS}. In Fig.~\ref{fig4} we show the rate
of this channel in the allowed parameter space, where the
model-independent $5\sigma$ discovery sensitivity is obtained by the
ATLAS collaboration for 30 fb$^{-1}$ integrated luminosity
\cite{ATLAS}. In the calculation of the signal rate, we used the
effective lagrangian method to incorporate the important SUSY
corrections. Our results show that for more than $99\%$ of the
survived samples, the rate is smaller than the discovery
sensitivity, which means that the LHC is unlikely to discover $H^+$.
Our results also indicate that the future XENON100 limits (in case
of nonobservation of DM) will further tighten the parameter space,
making the discovery of $H^+$ unlikely even with higher luminosity.
Note that for the charged Higgs lighter than top quark, the LHC
search can instead utilize top pair production with one top decay
into charged Higgs \cite{ATLAS}. Like the case of the heavy charged
Higgs boson, our results indicate that for more than $99\%$ of the
survived samples, the signal is below the $5\sigma$ discovery
sensitivity obtained by the ATLAS collaboration due to ${\rm Br}
(t\to H^+ b) < 10^{-2}$. The small branching ratio of $t\to H^+ b$
arises from the fact that $\tan \beta$ is around $10$ for $m_{H^+}
\leq 150$ GeV (see Fig.~\ref{fig1}) and for such a value of $\tan
\beta$ there is a strong cancellation between different terms in the
amplitude of this decay.

%%%%fig.5 %%%%%%%%%%%%%%%%%%%%%%%%%%%%%%%%%%%%%%%%%%%
\begin{figure}[htb]
\includegraphics[width=15cm]{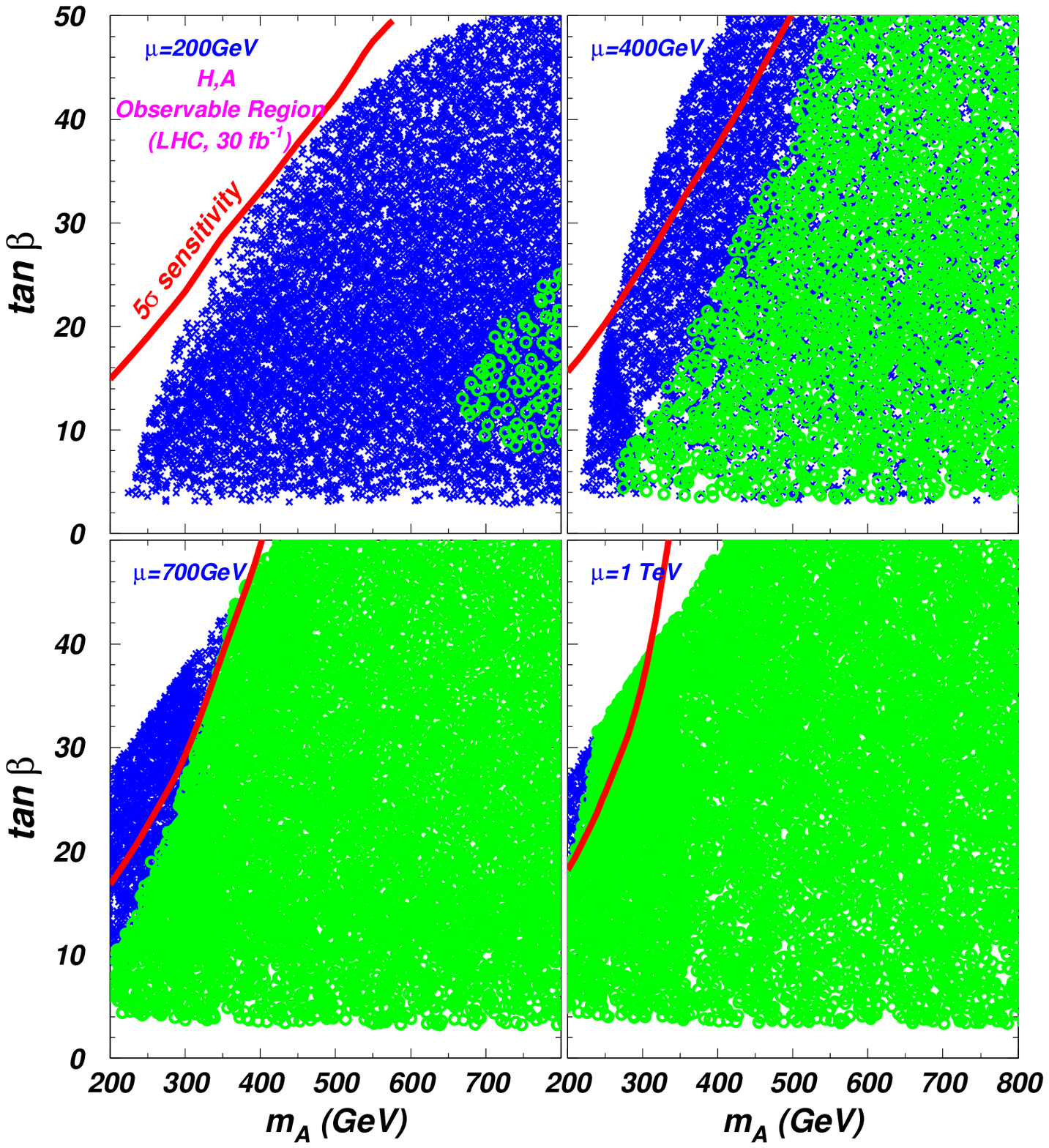}
\vspace{-0.8cm} \caption{Same as Fig.~\ref{fig3},
         but showing the LHC search sensitivity for the non-SM neutral Higgs bosons
         in the $m_h^{\rm max}$ scenario \cite{CMS-curve}.
Here the LHC sensitivity curve obtained by the CMS collaboration
\cite{CMS-curve} corresponds to the $5 \sigma$ discovery level,
while the exclusion limit from the dark matter
direct detection experiments is at $90\%$ C.L. } 
\label{fig5}
\end{figure}
%%%%%%%%%%%%%%%%%%%%%%%%%%%%%%%%%%%%%%%%%%%%%%%

Now we turn to the LHC search for the non-SM neutral Higgs boson $H$
and $A$, for which both the ATLAS and the CMS collaborations utilize
the channels $g g \to H(A)$ or $b\bar{b} H (A)$ with $H(A)$ decaying
to $\tau$ leptons \cite{ATLAS,CMS-curve,Horvat}. Unlike the charged
Higgs boson search, for which a model-independent discovery
sensitivity can be obtained, the analysis for the search of the
neutral Higgs bosons is performed in certain SUSY scenarios. Here we
consider the $m_h^{\rm max}$ scenario with the following fixed
parameters: $M_{\rm SUSY}=1$ TeV, $M_2=200$ GeV, $m_{\tilde{g}}=800$
GeV, and $X_t = A_t - \mu \cot \beta = 2$ TeV. In order to show the
$\mu$ dependence of the constraints, we choose several
representative values of $\mu$ and scan the rest free parameters in
the ranges: $ 1 \leq \tan \beta \leq 80 $, $ 80 {\rm~GeV} \leq m_A,
m_{\tilde{\ell}} \leq 0.8 {\rm~TeV} $ and $ 30 {\rm~GeV} \leq M_1
\leq 500 {\rm~GeV} $. In Fig.~\ref{fig5} we show the surviving
samples on $m_A$ versus $\tan\beta$ plane together with the LHC
discovery sensitivity for 30 fb$^{-1}$ integrated luminosity. This
sensitivity is obtained by the CMS collaboration with $H/A \to
\tau^+ \tau^- \to \mu +jets$ topology (semi-leptonic final states)
\cite{CMS-curve}, which is better than that obtained by the ATLAS
collaboration with $H/A \to \tau^+ \tau^- \to 2 \ell + 4 \nu$
topology (full leptonic final states) \cite{ATLAS,Horvat}. Note that
in Fig.~\ref{fig5} the LHC sensitivity curve corresponds to the $5
\sigma$ discovery level, while the exclusion
limit from the dark matter direct detection
experiments is at $90\%$ C.L.. 

A few comments are due regarding the results displayed in  Fig.~\ref{fig5}:
\begin{itemize}
\item[(1)] In getting these results
we used the package NMSSMTools (version 2.3.1) \cite{NMSSMTools}
which uses micrOMEGAs (version 2.2) \cite{Belanger}
for the calculation of the dark matter relic density.
But in our calculations we extended the package
by including more experimental constraints, such as the LEP
search for the Higgs bosons and $B^+ \to \tau^+ \nu_\tau$, so
our combined constraint on the parameter space is more stringent.

\item[(2)] The CDMS-II/XENON constraints are sensitive to the value of
$\mu$, i.e., as $\mu$ gets larger, the constraints become weak.
The reason for this behavior is that a larger $\mu$ will result in a
smaller Higgsino component in $\chi^0_1$ and hence suppress the
Higgs-$\chi^0_1$-$\chi^0_1$ coupling, which will weaken the
CDMS-II/XENON constraints.

\item[(3)] The LHC sensitivity for $\mu=200$ GeV is taken directly
from \cite{CMS-curve}, and for other values of $\mu$ the curves are
obtained by scaling the value of $\tan \beta$ so that the production
rate of the Higgs bosons is same as that for $\mu=200$ GeV. In doing
this we used the package FeynHiggs2.7.1 \cite{FeynHiggs} to
calculate the production rate. Note that the $\mu$ parameter affects
the production rate mainly by changing the $H(A) b\bar{b}$ coupling
through loop correction ($\Delta_b$) which is proportional to $\mu
\tan\beta/M_{SUSY}$ \cite{Carena}. So the shift of the LHC
sensitivity curve due to the variation of the $\mu$ value is not
negligible, as shown in Fig.~\ref{fig5}.

\item[(4)] Fig.\ref{fig5} shows that for $\mu=200$ GeV no
surviving samples can reach the observable level, while for larger
$\mu$ values a small fraction of surviving samples can lie within
the observable region. Numerically we checked that for $\mu=400
{\rm GeV}, 700 {\rm GeV}$, $1{\rm TeV}$, about $8\%$, $11\%$ and
$7\%$ of the surviving samples lie within the observable region,
respectively (for $\mu=1$ TeV about one third of these detectable
samples can even survive the future XENON100 limit). The reason
for this behavior is that for a large $\mu$, although the LHC
sensitivity curve is shifted upward, due to the much weakened
CDMS-II/XENON constraints, some surviving samples can have quite
large $\tan \beta$ values (as shown in Fig.\ref{fig3} and
Fig.\ref{fig5}) so that they can reach the LHC sensitivity.

\item[(5)] About the lower bound of $M_A$ as a function of $\tan \beta$,
since both the $\chi N$ scattering cross section and the
production rate of the Higgs boson are proportional to $\tan^2
\beta$ for large $\tan \beta$, one would naively expect the
lower bound curve runs in parallel with the LHC sensitivity curve.
As shown in Fig.\ref{fig5}, this is not the case because we
considered many experimental constraints and not all of them
scale as $\tan^2 \beta$.
\end{itemize}
\vspace{0.5cm}

{\em Conclusion:}
We have seen that if the MSSM is the true story, the current limits
from dark matter and collider experiments already strongly
constrain the parameter space, which has important implication
for the LHC searches for the non-SM SUSY Higgs bosons.
It turns out that in the currently allowed parameter space, the charged
Higgs boson is hardly accessible at the LHC for an integrated
luminosity of 30 fb$^{-1}$, while the neutral non-SM Higgs bosons
($H$,$A$) may be accessible in some allowed region characterized by
a large $\mu$.
The future XENON100 (6000 kg-days exposure) will significantly
tighten the parameter space in case of nonobservation of dark matter,
further shrinking the likelihood of discovering the non-SM Higgs bosons
at the LHC. So the interplay of the dark matter direct detection experiments
and the LHC Higgs searches will allow for a good test of the MSSM.

Finally we stress that we obtain the above conclusion by choosing a
small $f_{Ts}$ ($=0.02$). If we choose a large $f_{Ts}$, the
scattering rate will be larger so that the limits from the current
CDMS II/XENON will become more stringent. For example, for
$f_{Ts}=0.38$ taken in previous studies \cite{Ellis}, we find that
the current CDMS II/XENON constraints are comparable with the future
XENON100 (6000 kg-days) constraints. We also checked that if we
relax some assumptions in our scan, e.g., the grand unification
relation for the gaugino masses, our findings about the LHC Higgs
searches remain unchanged. Further, we noticed the controversy on
the XENON100 detection efficiency \cite{last}. Although our current
bounds are from CDMS II plus XENON100, the CDMS II plays the dominant
role. If we do not include the current XENON100 limits, our results
almost remain unchanged.
\vspace{0.2cm}

{\em Note added:~} After finishing our paper, we notice that the ATLAS
collaboration publishes an analysis for the search sensitivity of  
the neutral non-SM Higgs bosons via the semi-leptonic final states 
\cite{ATLAS-LH}, in which the obtained discovery sensitivity seems 
better than the CMS result shown in Fig.\ref{fig5}. 
According to this new ATLAS result, more surviving samples in Fig.\ref{fig5} 
will reach the observable region. So our conclusion about the observability 
of the neutral Higgs bosons will remain unchanged.
\vspace{0.2cm}

{\em Acknowledgment:~}
We thank Tao Han for discussions.
This work was supported in part by the Grant-in-Aid for Scientific
Research (No.~14046201) from Japan and by the NSFC
(Nos.~10821504, 10725526, 10635030) from China.

\end{document}